\documentclass{article}

\usepackage[nonatbib,final]{neurips_2020}

\usepackage[utf8]{inputenc} 
\usepackage[T1]{fontenc}    
\usepackage{hyperref}       
\usepackage{booktabs}       
\usepackage{amsfonts}       
\usepackage{nicefrac}       
\usepackage{microtype}      
\usepackage{subfigure}      
\usepackage{cite}
\usepackage[colorinlistoftodos]{todonotes}

\usepackage{amsmath}
\usepackage[squaren,Gray]{SIunits}
\usepackage{tikz}

\definecolor{redtpt}{rgb}{0.749,0.071,0.220}
\definecolor{edf1}{rgb}{0, 0.102, 0.439}

\usetikzlibrary{calc,quotes}

\title{Towards Optimal District Heating Temperature Control in China with Deep Reinforcement Learning}

\author{
Adrien Le Coz \\
EDF R\&D China Center\\
Beijing, China \\
\texttt{adrien.le-coz@edf.fr} \\
\And
Tahar Nabil \\
EDF R\&D China Center \\
Beijing, China \\
\texttt{tahar-t.nabil@edf.fr} \\
\And
Fran{\c{c}}ois Courtot \\
EDF R\&D China Center \\
Beijing, China \\
\texttt{francois.courtot@edf.fr} \\
}

\graphicspath{{figures/}}

\begin{document}

\maketitle

\begin{abstract}

Achieving efficiency gains in Chinese district heating networks, thereby reducing their carbon footprint, requires new optimal control methods going beyond current industry tools.
Focusing on the secondary network, we propose a data-driven deep reinforcement learning (DRL) approach to address this task.
We build a recurrent neural network, trained on simulated data, to predict the indoor temperatures.
This model is then used to train two DRL agents, with or without expert guidance, for the optimal control of the supply water temperature.
Our tests in a multi-apartment setting show that both agents can ensure a higher thermal comfort and at the same time a smaller energy cost, compared to an optimized baseline strategy.
\end{abstract}

\section{Introduction}

Space heating and cooling in buildings is well-known for representing a significant part of global CO\textsubscript{2} emissions.
For instance, district heating alone in China consumes more energy than the entire United Kingdom \cite{IEA2017}.
As of today, operating such industrial heating networks is prone to energy losses due to complex nonlinear building thermal behaviours.
Hence significant efficiency gains are possible resulting in a clear line of work to fight climate change:
develop and implement advanced control strategies for an optimal operation.
This is all the more relevant in China where 87\% of heat production is either from coal or oil \cite{IEA2018}.

\begin{figure}
\centering
\begin{tikzpicture}
\begin{scope}[scale=0.65]
\draw[dashed] (.625,-0.4) -- (.625,5);
\draw[rounded corners=4] (0,0.25) rectangle (1.25,1.75);
\draw[] (0.3125,1.75) -- (0.3125,0.25);
\draw[] (0.625,1.75) -- (0.625,0.25);
\draw[] (0.9375,1.75) -- (0.9375,0.25);
\draw[color=redtpt,rounded corners, very thick] (-2.25,1.5) -- (0.16,1.5) -- (0.16,1);
\draw[color=edf1,rounded corners,very thick] (0.16,1) -- (0.16,0.5) -- (-2.25,0.5);
\draw[color=edf1,rounded corners, very thick] (11,0.5) -- (1.09,0.5) -- (1.09,1);
\draw[color=redtpt,rounded corners,very thick] (1.09,1) -- (1.09,1.5) -- (11,1.5);
\draw[color=redtpt, very thick,->] (-2.25,1.5) -- (-0.9,1.5);
\draw[color=edf1, very thick,->] (-0.5,0.5) -- (-0.9,0.5);
\draw[color=redtpt, very thick,->] (1.5,1.5) -- (2.05,1.5);
\draw[color=edf1, very thick,->] (5,0.5) -- (2.05,0.5);

\draw[very thick,color=redtpt] (11.5,1.5) node {$\cdots$};
\draw[very thick,color=edf1] (11.5,0.5) node {$\cdots$};
\draw[very thick,color=redtpt] (-3.,1.5) node {$\cdots$};
\draw[very thick,color=edf1] (-3.,0.5) node {$\cdots$};

\draw[->,>=latex] (-4,1) -- node[above] {\scriptsize \emph{towards heat}} node[below] {\scriptsize \emph{generation plant}} (-6.25,1);

\draw[fill=white] (-.7,1.65) -- (-.3,1.35) -- (-.3,1.65) -- (-.7,1.35) --cycle;
\draw[] (-.6,1.65) rectangle (-.4,1.85);
\draw[] (-0.5,1.5) -- (-0.5,1.65);

\node[] (p1) at (1.75,1.5) {\textbullet};
\node[] (p2) at (1.75,0.5) {\textbullet};
\node[] (t1) at (1.5,2.5) {\scriptsize$T_{s}$};
\node[] (t2) at (1.5,-0.25) {\scriptsize$T_{r}$};
\draw[] (1.75,1.5) -- (t1);
\draw[] (1.75,0.5) -- (t2);
\node[] (hx) at (-.65,-.25) {\scriptsize Heat Exchanger};
\draw[] (hx) -- (0,0.25);

\draw[fill=white] (2.75,0.5) circle (0.2);
\draw[] (2.75,0.5)++(45:0.2) -- (2.55,0.5);
\draw[] (2.75,0.5)++(-45:0.2) -- (2.55,0.5);
\node[] (m) at (2.75,0.1) {\scriptsize$\dot{m}$};

\node[draw,rectangle,rounded corners=7.pt] (P) at (-.55,3.75) {\scriptsize\emph{Primary}};
\node[draw,rectangle,rounded corners=7.pt] (S) at (1.85,3.75) {\scriptsize\emph{Secondary}};

\draw[] (3.5,2.5) rectangle (6.25,6.25);
\draw[dashed] (3.5,4.75) -- (6.25,4.75);
\draw[] (4,2.5) rectangle (4.5,3);

\draw[] (7,2.5) rectangle (9.75,6.25);
\draw[dashed] (7,4.75) -- (9.75,4.75);
\draw[] (7.5,2.5) rectangle (8,3);

\begin{scope}[shift={(3.875,6.5)},scale=1]
\draw[rounded corners=2] (1,-3.2) -- (2,-3.2) -- (2,-3) -- (0,-3) -- (0,-2.8) -- (2,-2.8) -- (2,-2.6) -- (0,-2.6) -- (0,-2.4) -- (2,-2.4) -- (2,-2.2) -- (-0.1,-2.2) (1,-1.6) -- (2,-1.6) -- (2,-1.4) -- (0,-1.4) -- (0,-1.2) -- (2,-1.2) -- (2,-1) -- (0,-1) -- (0,-0.8) -- (2,-0.8) -- (2,-.6) -- (-0.1,-.6);
\end{scope}

\begin{scope}[shift={(7.375,6.5)},scale=1]
\draw[rounded corners=2] (1,-3.2) -- (2,-3.2) -- (2,-3) -- (0,-3) -- (0,-2.8) -- (2,-2.8) -- (2,-2.6) -- (0,-2.6) -- (0,-2.4) -- (2,-2.4) -- (2,-2.2) -- (-0.1,-2.2) (1,-1.6) -- (2,-1.6) -- (2,-1.4) -- (0,-1.4) -- (0,-1.2) -- (2,-1.2) -- (2,-1) -- (0,-1) -- (0,-0.8) -- (2,-0.8) -- (2,-.6) -- (-0.1,-.6);
\end{scope}

\draw[color=redtpt,very thick] (3.75,1.5) node {\textbullet} -- (3.75,4.3) node {\footnotesize\textbullet} -- (3.75,5.9) node {\footnotesize\textbullet};

\draw[color=edf1,very thick] (4.9,0.5) node {\textbullet} -- (4.9,3.3) node {\footnotesize\textbullet} -- (4.9,4.9) node {\footnotesize\textbullet};

\draw[color=redtpt,very thick] (7.25,1.5) node {\textbullet}-- (7.25,4.3) node {\footnotesize\textbullet} -- (7.25,5.9) node {\footnotesize\textbullet};

\draw[color=edf1,very thick] (8.4,0.5) node {\textbullet}  -- (8.4,3.3) node {\footnotesize\textbullet} -- (8.4,4.9) node {\footnotesize\textbullet};

\draw[color=redtpt, very thick,->] (4.5,1.5) -- (5.5,1.5);
\draw[color=edf1, very thick,->] (8,0.5) -- (6.65,0.5);
\end{scope}
\end{tikzpicture}
\caption{A district heating network.}
\label{fig:dh}
\end{figure}
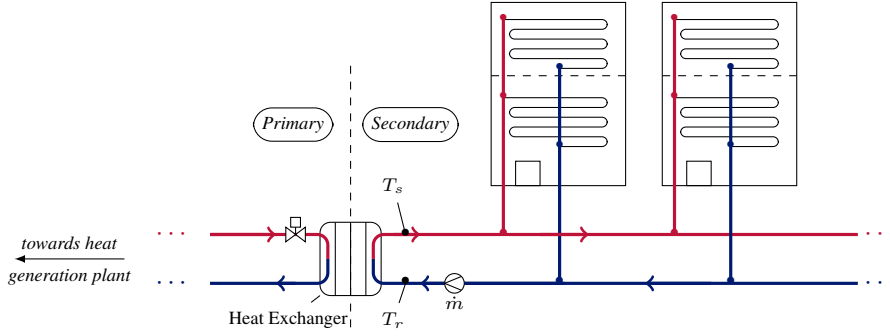

In a typical Chinese district heating system,
heat is produced in a central location and conveyed towards substations and onwards for distribution to customers via a heat exchanger and a network of insulated pipes.
The distribution network is organized in a feed-and-return line and contains two parts, the primary and secondary networks as shown in Figure \ref*{fig:dh}.
The secondary network contains temperature sensors $T_s$, for the fluid exiting the substation towards the feed line, and $T_r$ for the fluid entering the heat exchanger back from the return line.
As is the norm in China and unlike some contributions on optimal district heating such as \cite{giraud2017optimal,cox2019real,saloux2020optimal}, we assume that the network operator buys heat at a constant price and from a unique third-party producer.
We focus thus on the control of the secondary indoor temperatures:
heat must be delivered to each apartment on the feed line to ensure that every indoor temperature is within an admissible range.
In particular, any temperature above the upper bound results in both a waste of energy and an economic loss for the utility - most often, heating fees in China are a flat cost per square meter, regardless of the actual energy consumption.
However, it is not possible to individually control the thermal behaviour of every apartment.
Instead, an operator should control the indoor temperatures with two commands located inside the substation:
(i) the supply temperature $T_{s}$, by acting on a control valve on the primary side, and
(ii) the flow rate of the fluid, thanks to a pump at the inlet bound of the secondary side of the heat exchanger.

The state-of-the-art industrial control strategies rely on a relationship, called the water curve, which is tuned by an expert, between $T_{s}$ and the outdoor temperature $T_{o}$.
This paper investigates how to improve this strategy to ensure both higher thermal comfort and smaller energy and operation cost.

\section{Approach}

\subsection{Control strategy}

We apply a Reinforcement Learning (RL) paradigm \cite{sutton1998introduction},
where an \emph{agent} learns a control strategy (\emph{policy}) by interacting with the environment - here, the set of rooms heated by the network.
The problem is modelled as a Markov Decision Process: the agent receives an observation of the \emph{state} of the environment, chooses an \emph{action} and receives as a result a \emph{reward} from the environment.
The best control strategy maximizes the expected cumulative discounted reward over the lifetime of the agent.
Learning such a policy requires first to derive a model of the environment, to predict the indoor temperatures from the commands and the weather conditions.
This model is described in Section \ref*{ssec:identification}.

At time $t$, the state is a vector $s_t$ containing the outdoor temperature $T_{o,t}$, supply water temperature $T_{s,t}$, time of the day and indoor temperatures $T_{in,t}^{(j)}$ for every room $j\in\{1,\dots,N\}$ in the network.
$s_t$ contains both present and past $n$ measurements of these quantities.
At an hourly time step, a history of 24 hours is used to form $s_t$.
At that same hourly time step, the agent is asked to select an action $a_t$.
The flow rate being kept constant, the action is restricted to the supply temperature $T_{s,t}$.
Two discrete action spaces, with $T_s(\celsius)\in\{20,21,\dots,50\}$, are considered.
Agent 1 is the standard strategy while Agent 2 is a finetuning of the baseline control strategy (cf Section \ref{ssec:baseline}):
\begin{enumerate}
\item Agent 1: to enforce the smoothness of the control signal, the action is limited to the increments $a_t = T_{s,t}- T_{s,t-1}$ where $a_t\in\mathcal{A}:=\{0,\,\pm0.5,\,\pm1,\,\pm1.5,\,\dots,\pm3\}$.
\item Agent 2: the discrete action is the difference
$a_t = T_{s,t}- T_{s,t}^{b}$ where
$a_t\in\mathcal{A}$ and $T_s^b$ is the estimated baseline supply temperature.
\end{enumerate}
Finally, the agent selects the action in order to maximize the expected cumulative discounted reward function $R=\sum_{t=0}^{T}\gamma^tr(s_{t},a_{t})$ over $T$ time steps in the heating season.
In the sequel, the discount factor is set to $\gamma=0.9$, which corresponds to an agent that adapts its behaviour to the expected reward for the next 30 hours.
The reward $r$ penalizes deviations from a target temperature $\mathcal{T}$:
\begin{align}\label{eq:reward}
r(a_t,s_t) = -\sum_{j=1}^N |T_{in,t}^{(j)}-\mathcal{T}_t^{(j)}|.
\end{align}

We use Deep Reinforcement Learning (DRL), to train the different agents.
DRL has proven to be a successful algorithm in various domains such as games, robotics or demand response \cite{li2017deep}.
In particular, we train Deep Q-Networks (DQNs, \cite{mnih2013playing,mnih2015human}). 
For each training episode, a weather file is randomly chosen from a set of 7 cities in China to avoid overfitting the local climate, and an entire heating season is simulated.
The weather measurements for testing the agents come from an eighth city, Yuncheng.
Some statistics summarizing the climate in these cities are gathered in Appendix \ref*{app:additional}.

\subsection{Model identification} \label{ssec:identification}

Consider a six-story building with three apartments per level - facing either the Eastern, Southern or Western direction.
Heat is provided from a district heating substation and supplied to the apartments through a high-inertia radiant heating system.
The detailed building dynamics are simulated from an in-house model developed using Dymola, a commercial physics-based modelling and simulation tool, and calibrated against real operation data to represent a variety of indoor temperatures found in a district rather than in a single building.
Seven apartments are thus left empty, to represent thermal losses and eleven indoor temperatures are simulated at an hourly time-step from two sets of inputs: the weather conditions (hour of the day, outdoor temperature $T_o$, solar flux and angles) and the commands (supply water temperature $T_s$ and mass flow rate $\dot{m}$).
Another output of the model is the water return temperature $T_r$.
The model simulates a heating season from mid-December to mid-March for a total of 83 days (2002 hours).

The detailed physical model is then used as a data generator process to train a statistical model predicting the indoor and return temperatures.
This model is a recurrent neural network (RNN, \cite{williams1989learning}) with two layers of 32 Long-Short Term Memory units \cite{hochreiter1997long}, a class of neural network able to catch long-term dependencies in sequential data.
Due to the high inertia of the system, the inputs are the same as for the detailed model, except that a sequence of 120 time-steps ($t-119$ to $t$) is used to make predictions at time $t$.
1,373 series are generated, of which 80\% are for training, 10\% for validation and 10\% for testing.
Each simulation has a random command and a weather file chosen from a random location in China (see Appendix \ref*{app:additional}).
Implemented in Python with \texttt{keras-tensorflow} \cite{chollet2015keras}, the model achieves a mean absolute test error of 0.110\celsius{} with a standard deviation of 0.132\celsius{}.

Switching from expert model to RNN reduces the prediction time - about 30 times faster on CPU - which enables an efficient training of the agent.
Although a challenge in practice, RNN networks can also be fine-tuned to on-site measurements with transfer learning, ensuring thus a higher scalability.

\subsection{Baseline model and metrics} \label{ssec:baseline}

The baseline model is a linear water curve $T_s^b = \alpha+\beta T_o$, $\alpha,\,\beta$ being found by minimizing $\sum_t -r_t$, $r_t$ as in \eqref{eq:reward}, with Particle Swarm Optimization \cite{eberhart1995new}.
The optimal linear water curve is an advanced industry strategy, also called reactive control \cite{saloux2020optimal}.
Besides, we implement and tune a PID controller, a frequently used strategy for temperature control \cite{zhang2019whole}.
All computations are carried out in Python.

The different policies are compared in terms of (i) thermal comfort, i.e. deviation from target temperature and stability of indoor temperatures, (ii) energy cost and (iii) estimated CO\textsubscript{2} cost.
The energy cost is computed from the temperature difference $T_s-T_r$ with $T_r$ predicted by the RNN model.
The CO\textsubscript{2} cost is the amount of CO\textsubscript{2} emissions for one square meter of heated surface and per heating season, under the standard industry assumption that it requires 80 kWh\per\squaremetre{} of heat from a coal-fired co-generation power plant, a wide-spread technology in China.
Detailed computations are presented in Appendix \ref{app:additional}.

\section{Results and Discussion}

\begin{figure}
\centering
\subfigure
{
\includegraphics[width=0.45\textwidth]{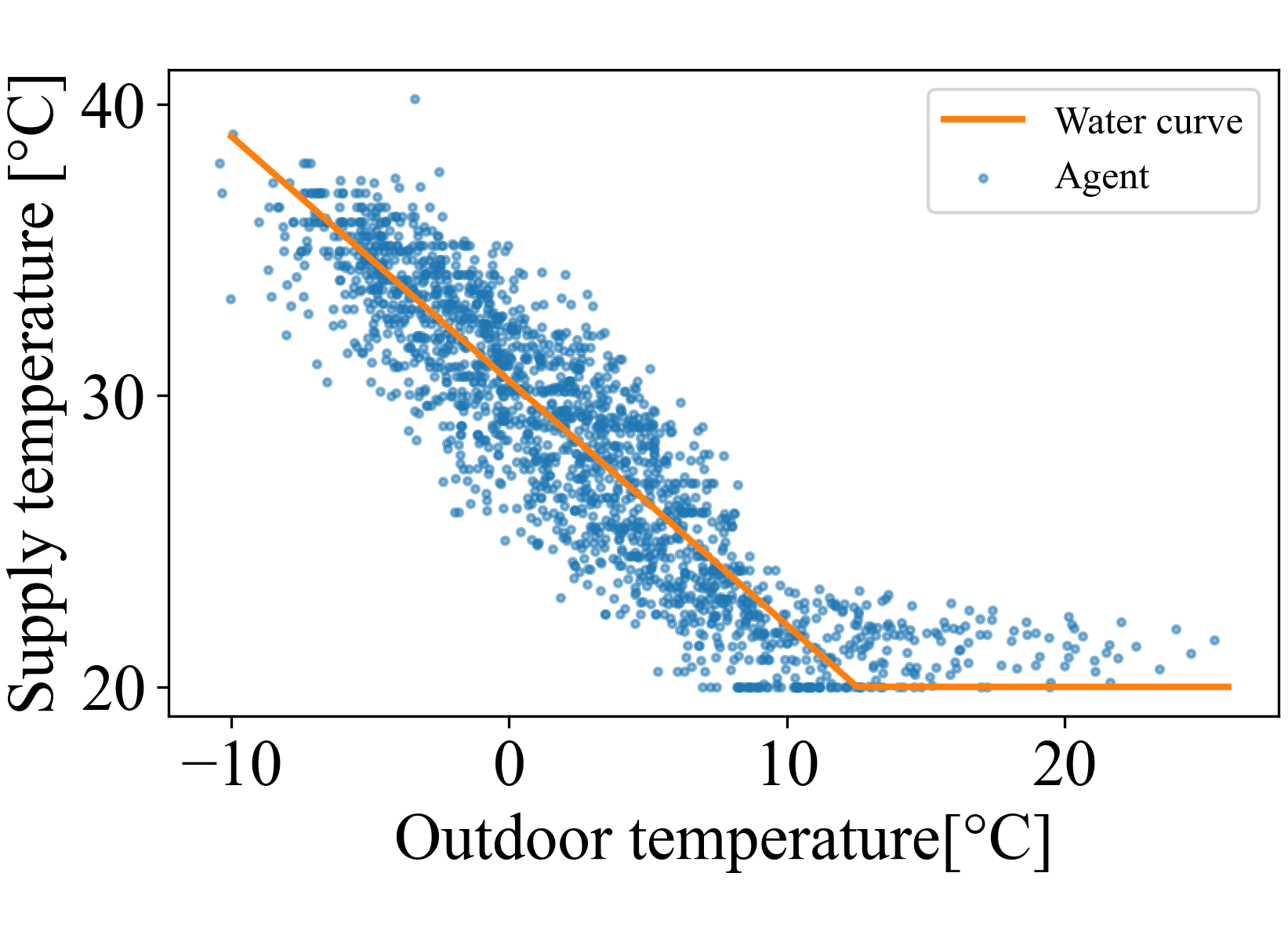}
}
\subfigure
{
\includegraphics[width=0.45\textwidth]{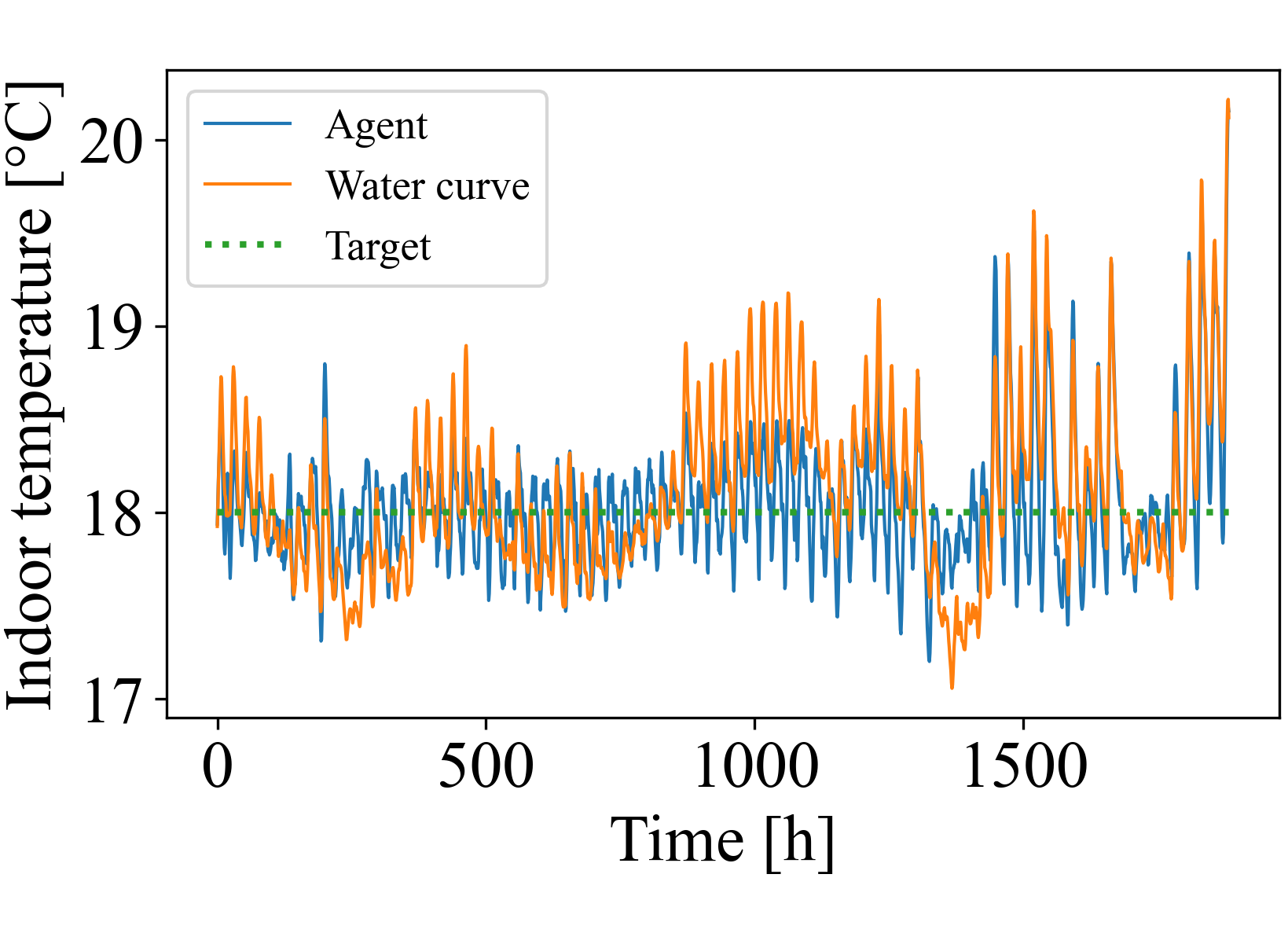}
}

\caption{Agent 1 and Baseline performances for one apartment.
(left) Control strategies $T_{s}$ vs $T_{o}$ and
(right) trajectories of $T_{in}$.
Constant target at 18\celsius.
Best viewed in colors.
}
\label{fig:res-tin-one-apt}
\end{figure}

We consider first the control of a single apartment with constant target $\mathcal{T}\equiv18\celsius$.
It can be seen from Figure \ref*{fig:res-tin-one-apt} that Agent 1 uses the extra degree of freedom on $T_s$ to maintain a more stable temperature, closer to the target, hence a higher thermal comfort.
Besides, the energy cost of Agent 1 is 2.73\% smaller than the baseline. 
Similar conclusions apply to Agent 2 (see Appendix \ref*{app:additional}, Figure \ref*{fig:res-tin-one-apt2}).

Next, Table \ref*{tab:results} summarizes the metrics for controlling the 11 indoor temperatures at the same time, with $\mathcal{T}\equiv 18$\celsius.
The Agents achieve a better control of indoor temperatures, combined with energy and CO\textsubscript{2} savings:
controlling a heating network serving one million square meters (about 20,000 customers) with Agent 2 would save about 495 tons of CO\textsubscript{2} per season.
Using the expert baseline strategy to guide the actions of the agent is also beneficial, as illustrated by the improved performances of Agent 2.
This might help increasing the acceptability of reinforcement learning for the network operator, since in this case the action is always at most 3\celsius{} different from the expert choice.
Finally, the PID controller performs slightly better than the baseline, but not as well as the Agents.

It can be noted that the environment contains little flexibility - a unique heat source at fixed price, no storage units - whereas advanced controls are designed to unlock it \cite{vandermeulen2018controlling,giraud2017optimal}.
Hence, robust heuristics like the optimal water curve are here difficult to improve further.
Yet, flexibility can be achieved by modifying the agent's reward  to allow more refined dynamic control targets, e.g. with a two-level target temperature $\mathcal{T}=17\celsius$ (night) and $\mathcal{T}=18\celsius$ (day), Agent 1 achieves 6.6\% energy savings.

\begin{table}[b]
\centering
\caption{
Performances of the control strategies in the multi-apartment setting, for $\mathcal{T}\equiv18\celsius$.
MAE: mean absolute error, \emph{std}: standard deviation.
Best performance is emphasized in \textbf{bold}.
}
\begin{tabular}{lcccc}
\toprule
& \textbf{MAE $T_{in}$} & \textbf{std $T_{in}$} & \textbf{Energy gain} & 
\textbf{CO\textsubscript{2} saved} \\
& \small  (\celsius) & \small (\celsius) & \small (\%) & \small (\gram/\squaremetre) \\
\textbf{Baseline} & 0.599 & 0.755 & 0 & 0 \\
\textbf{PID} & 0.584 & 0.742 & 0.95 & 215 \\
\textbf{Agent 1} & 0.549 & 0.699 & 2.15 & 486 \\
\textbf{Agent 2} & \textbf{0.545} & \textbf{0.692} & \textbf{2.19} & \textbf{495}\\
\bottomrule
\end{tabular}
\label{tab:results}
\end{table}

The good performances of the agents are in line with several other recently published papers applying reinforcement learning to controlling an indoor temperature, with smaller energy gains due to the lack of flexibility of the environment and an optimized baseline.
Besides, our approach differs from these contributions in some aspects.
First of all, whereas most studies focus on the thermal behaviour of a unique building, e.g. \cite{wei2017deep,nagy2018deep,zhang2018deep,jia2019advanced}, our system is a whole substation, 
with the 11 apartments originally tuned to represent the variety found in a district.
District heating system are studied in \cite{giraud2017optimal,zhang2019flow,saloux2020optimal,cox2019real} with promising results in terms of control, but with no analysis of the effect of their policies on indoor temperatures.
Secondly, these references assume a simple rule-based strategy for baseline, or a manual control \cite{zhang2019whole}.
For district heating however, using water curves for setting the supply temperature is common practice, with parameters tuned by hand from a set of a dozen indoor temperature sensors representative of the district thermal behaviour.
By optimizing these parameters, our baseline strategy is thus an improvement compared to the industry standards, in order to better assess the potential gain due to reinforcement learning.

Another key feature of applying reinforcement learning to real-world problems is the design of the reward function.
Most references build the reward as an explicit trade-off between energy cost and thermal comfort with careful weighting of the two contributions, see e.g. \cite{wei2017deep,zhang2018deep,zhang2019whole,nagy2018deep}.
On the contrary, our approach was to define a reward that depends solely on the target temperature specified by the contract between utility and customers, and to evaluate whether the agents can lower the energy cost as a side effect.
Indeed, when adding an energy cost in this low flexibility environment, the Agents maintain a mean indoor temperature constant at the lowest possible level; this effect can also be achieved by the baseline by lowering the target temperature.
Moreover, we find in our experiments that the reward function \eqref{eq:reward} is more stable and robust to different weather conditions, while still maintaining an advantage in terms of both energy cost and thermal comfort.

Nevertheless, our results suggest that deep reinforcement learning, by understanding the dynamics of the system, is a suitable tool for controlling district heating networks, maintaining thermal comfort while reducing energy cost.
In order to be applied to an actual network, the first step is to deploy onsite outdoor and indoor temperature sensors.
Next, either the RNN model or a lightweight statistical model (e.g. equivalent RC electrical networks) is finetuned on the operation data.
Based e.g. on a cloud infrastructure to store the measurements, the agents, whether they are DQN or more recent agents such as DDPG \cite{lillicrap2015continuous}, can finally be deployed for controlling the substation \cite{zhang2019flow}.

\newpage
\small

\bibliographystyle{IEEEtran}
\bibliography{IEEEabrv,main}

\newpage

\appendix

\section{Supplementary material} \label{app:additional}

\normalsize

\subsection{Weather information}

The weather files used in our numerical experiments come from eight cities in China.
Yuncheng's climate is used only for testing the different control strategies, while measurements from the seven other cities are used for training only.
Table \ref*{tab:weather} summarizes their climate in terms of indoor temperature and solar radiation.

\begin{table}[h]
\centering
\caption{Measured weather conditions averaged over one heating season.
$T_o$ (\celsius): outdoor temperature, $\max \phi$ (W.h/m$^2$): daily maximum global horizontal solar flux.
\emph{std:} standard deviation.
}
\resizebox{\textwidth}{!}{
\begin{tabular}{ccccccccc}
\toprule
 & \textbf{Beijing} & \textbf{Chengdu} & \textbf{Harbin}  & \textbf{Shenyang} & \textbf{Shijiazhuang} & \textbf{Xian} & \textbf{Yuncheng} & \textbf{Zhengzhou} \\
\midrule
mean $T_o$  & 6.92 & 12.36 & -3.37 & 1.87 & 8.27 & 8.40 & 2.53 & 8. 81 \\ 
(std $T_o$) & (7.58) & (4.78) & (10.01) & (1.87) & (7.20) & (6.67) & (5.81) & (6.69) \\
max $\phi$ & 634 & 374 & 533 & 582 & 555 & 510 & 631  & 580 \\
(std max $\phi$) & (186) & (261) & (167) & (198) & (211) & (226) & (315) & (230) \\
\bottomrule
\end{tabular}
}
\label{tab:weather}
\end{table}

\subsection{Metrics computation}

The energy cost is computed as follows.
With $T_s$ (respectively $T_r$) denoting the secondary supply (respectively return) water temperature, the heat transferred from the primary to the secondary side through the heat exchanger is:
$$
Q = \dot{m}\cdot c_p\cdot(T_s-T_r),
$$
where $\dot{m}$ is the water flow rate in \kilogram\per\second, $c_p=4180$~\joule\per\kilogram\per\kelvin{} and $Q$ is the heat duty (\watt).

\begin{figure}
\centering
\includegraphics[width=0.6\textwidth]{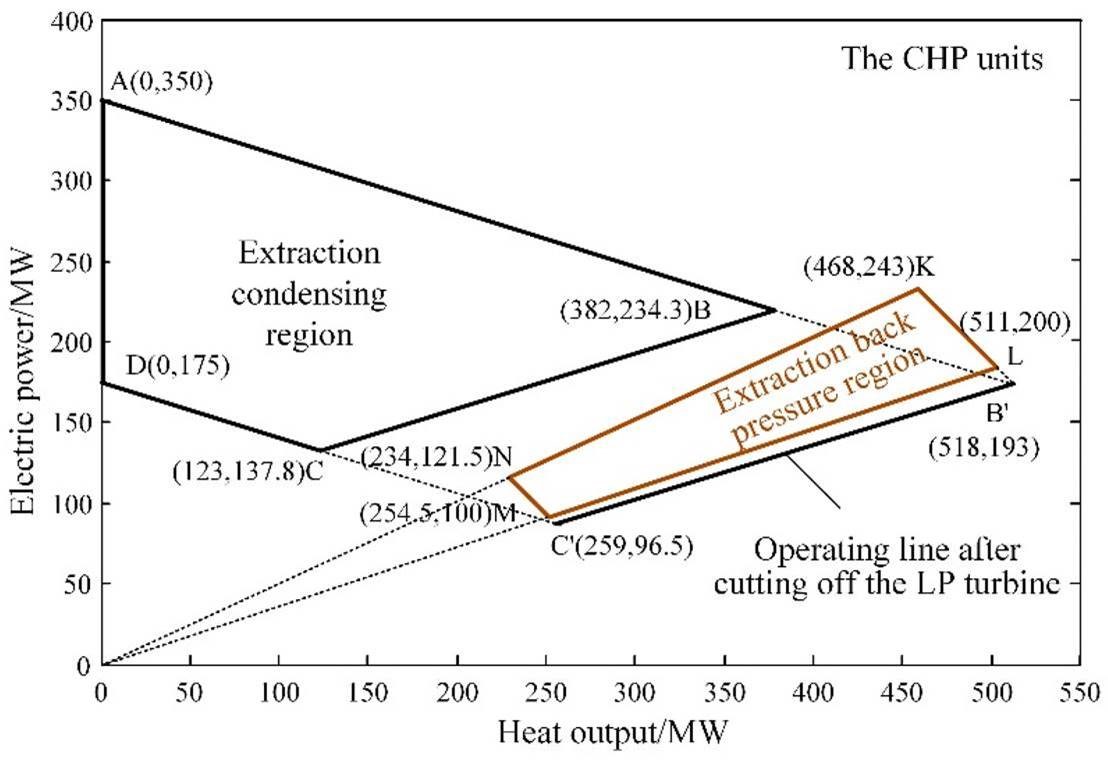}
\caption{Feasible operating regions for Combined Heat and Power Units. Source: \cite{zhang2020research}.}
\label{fig:coge}
\end{figure}

To estimate the CO\textsubscript{2} cost, we assume that heat is produced by a coal-fired co-generation power plant.
Under this technology, we make the assumption, standard in industry, that the primary heat consumption is 80~\kilo\watt\hour\per\squaremetre{}.
If a control policy saves $p\%$ energy, this corresponds thus to $80\cdot p$\kilo\watt\hour\per\squaremetre{} of saved primary heat.
Using the feasible operating region of combined heat and power plants shown in Figure \ref{fig:coge}, reducing the heat consumption by $80\cdot p$ increases the electricity generation by $80\cdot p\cdot \Delta$, where $\Delta = \left|\frac{350-234}{0-382}\right|$.
Finally, we assume than producing one more kilowatt-hour of electricity without increasing the coal consumption saves 930~\gram{} of CO\textsubscript{2}.
Hence the estimation of total amount of saved CO\textsubscript{2} emissions:
$$
930\times\left|\frac{350-234}{0-382}\right|\times 80 \times p \quad(\gram\text{CO}_2\per\squaremetre)
$$

\subsection{Additional figures and DQN hyperparameters}

The DQN has a learning rate set to 0.001,
buffer size to 1000000,
batch size to 32,
300 training episodes (1883 time steps per episode, after initialization during 119 steps); the initial (respectively final) value of random action probability is 1.0 (resp. 0.1), fraction of entire training period over which the exploration rate is annealed is 0.8,
the target network is updated every 200 steps.
The Q-network has two layers of 64 neurons each.

\begin{figure}[h]
\centering
\subfigure
{
\includegraphics[width=0.45\textwidth]{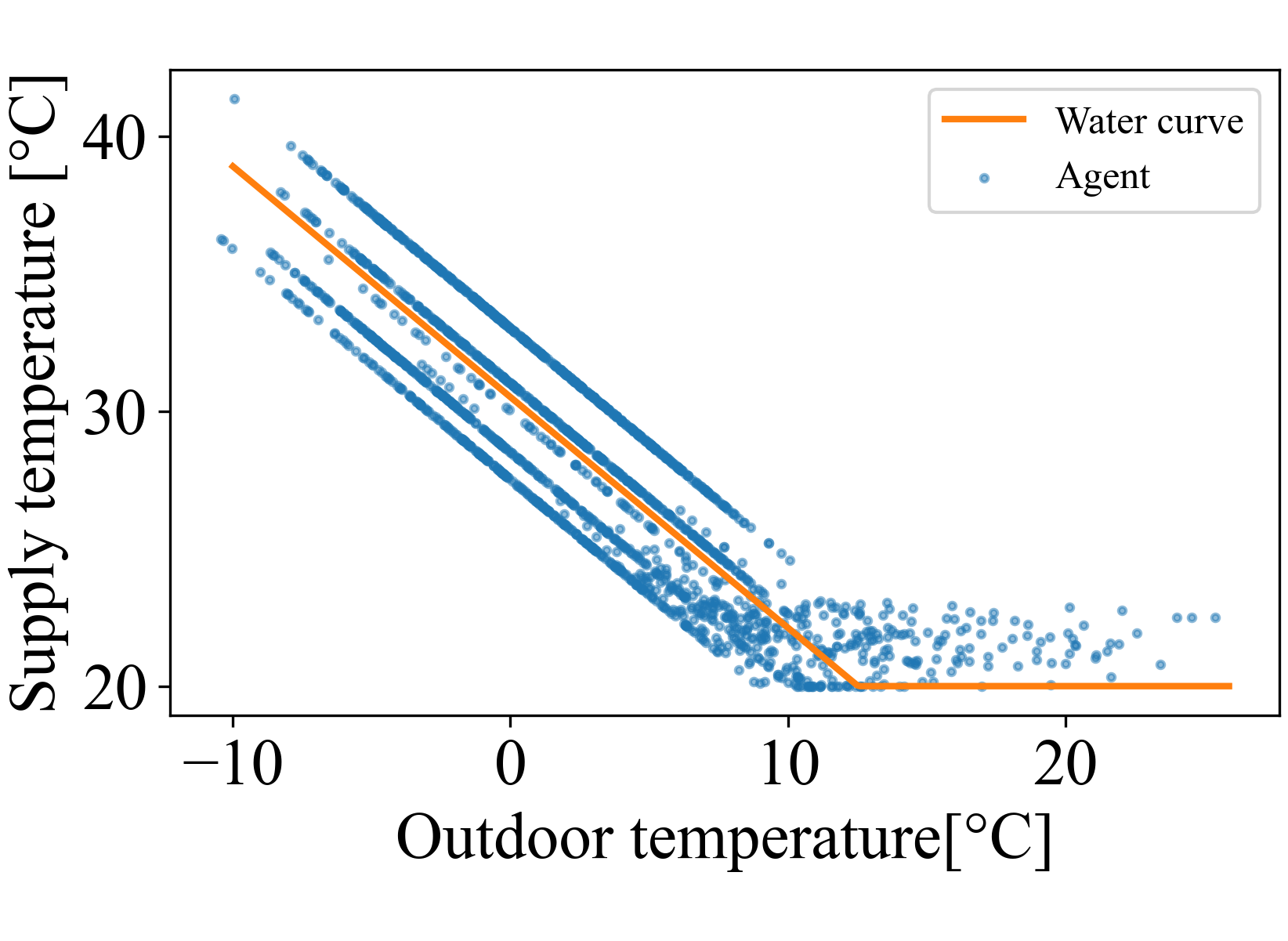}
}
\subfigure
{
\includegraphics[width=0.45\textwidth]{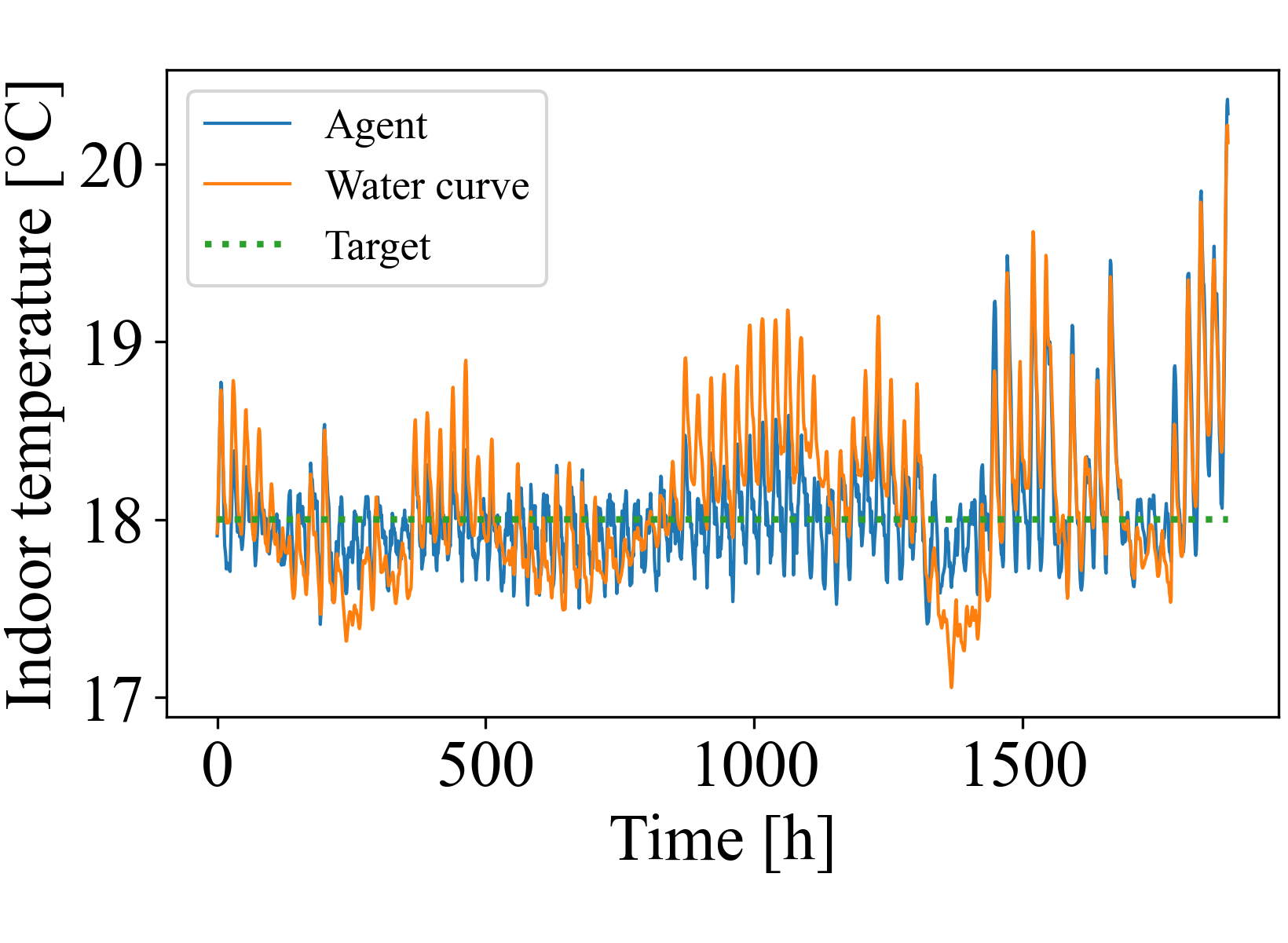}
}

\caption{Results for a single apartment, from Agent 2 and baseline strategy.
(left) Control strategy $T_{s}$ vs $T_{o}$ and
(right) performance in terms of $T_{in}$, through the heating season.
Best viewed in colors.
}
\label{fig:res-tin-one-apt2}
\end{figure}

The results for one apartment controlled by Agent 2 are displayed in Figure \ref*{fig:res-tin-one-apt2}.

\begin{figure}
\centering
\subfigure
{
\includegraphics[width=0.45\textwidth]{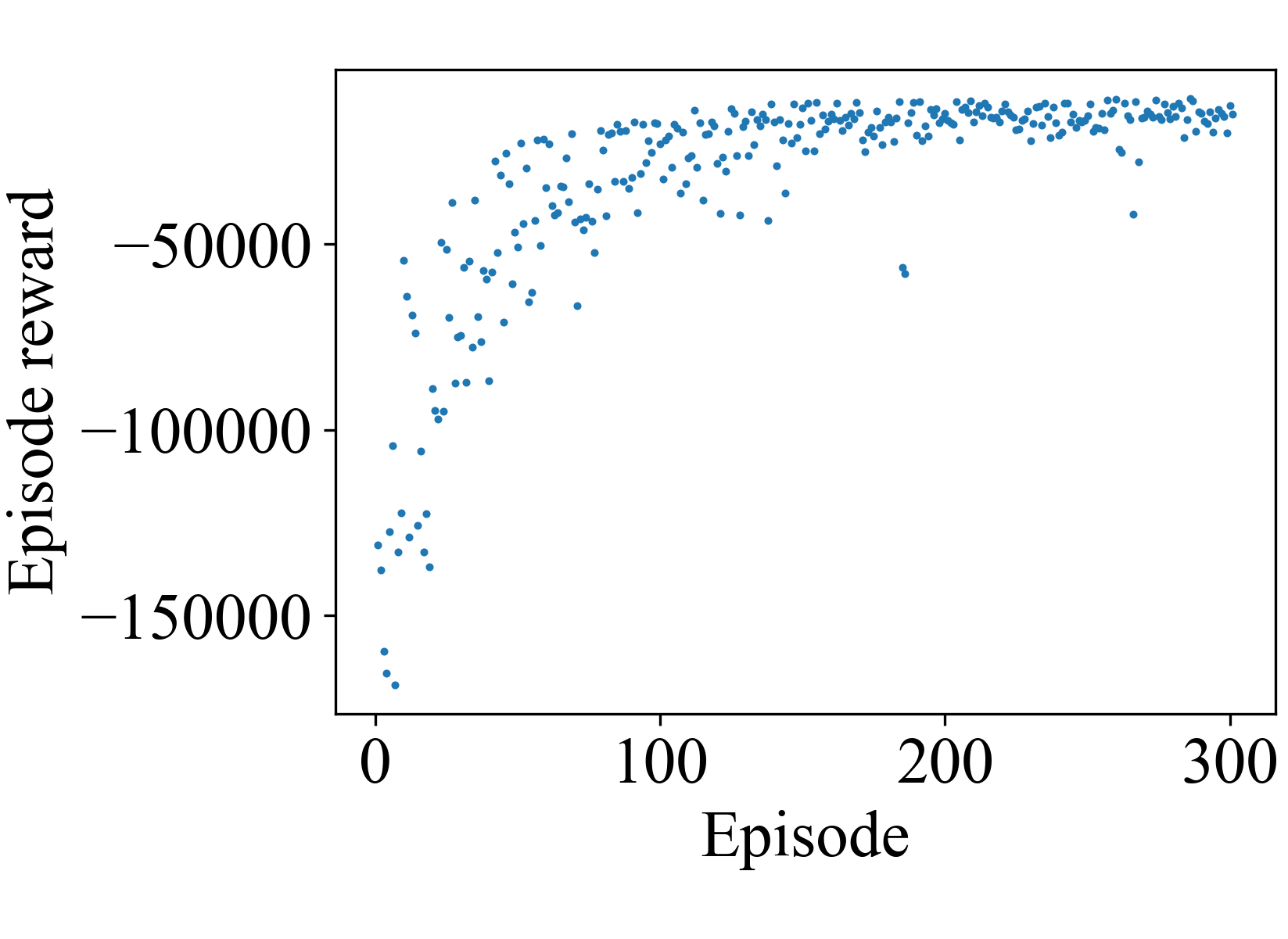}
}
\subfigure
{
\includegraphics[width=0.45\textwidth]{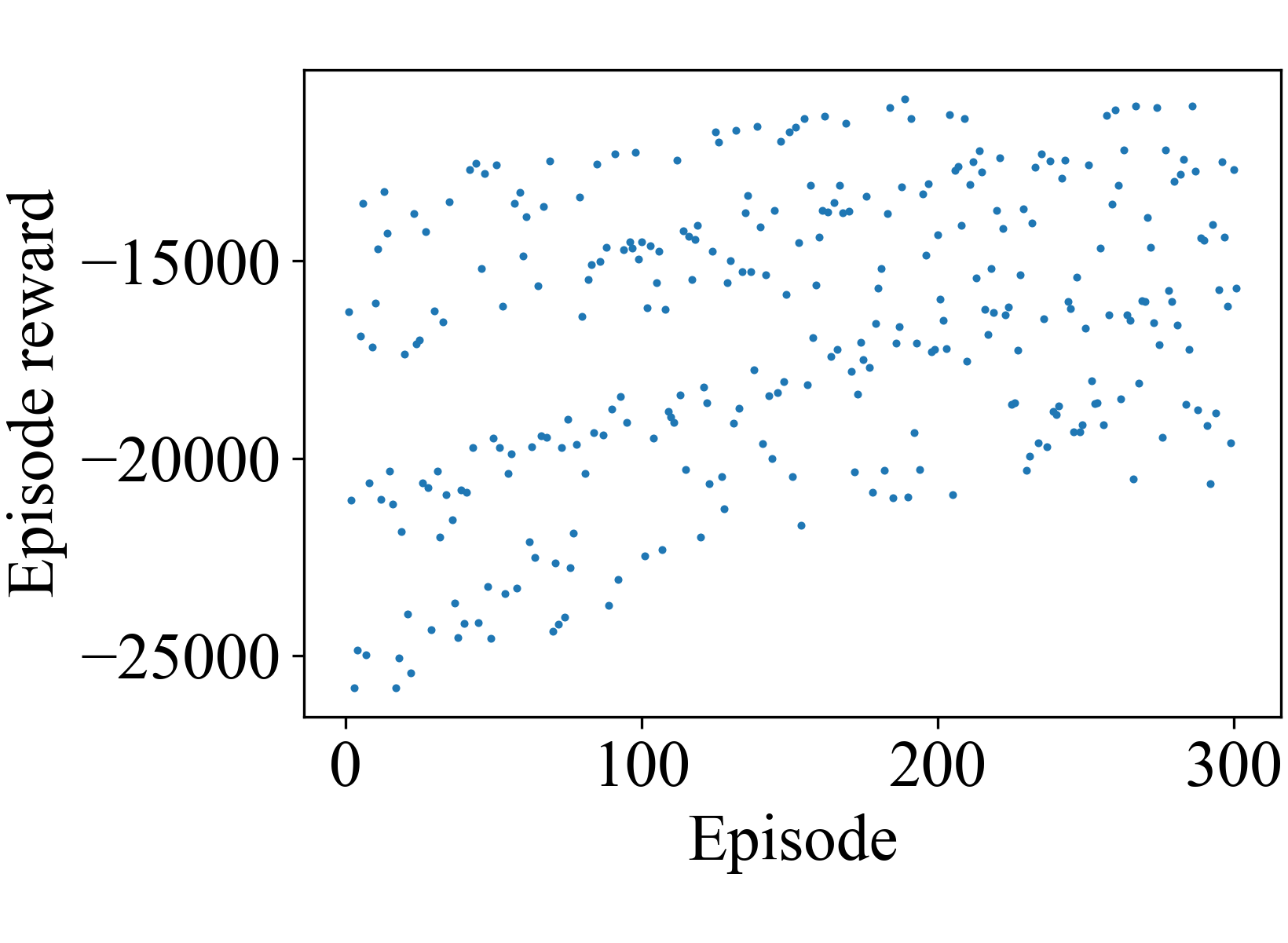}
}
\caption{Rewards during training of Agents 1 (left) and 2 (right) for controlling all apartments.
}
\label{fig:rewards}
\end{figure}

\begin{figure}
\centering
\subfigure[]
{
\includegraphics[width=0.45\textwidth]{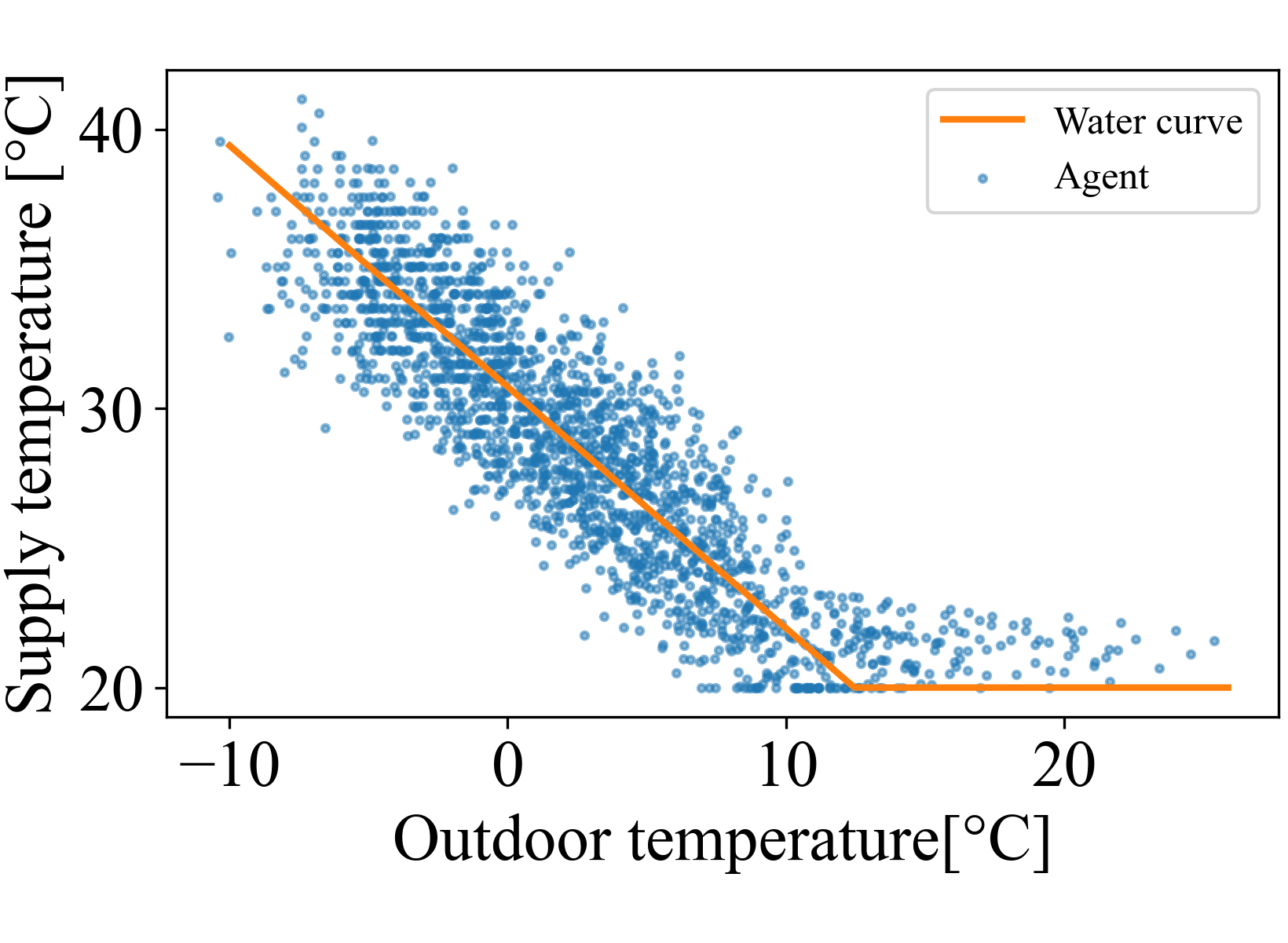}
}
\subfigure[]
{
\includegraphics[width=0.45\textwidth]{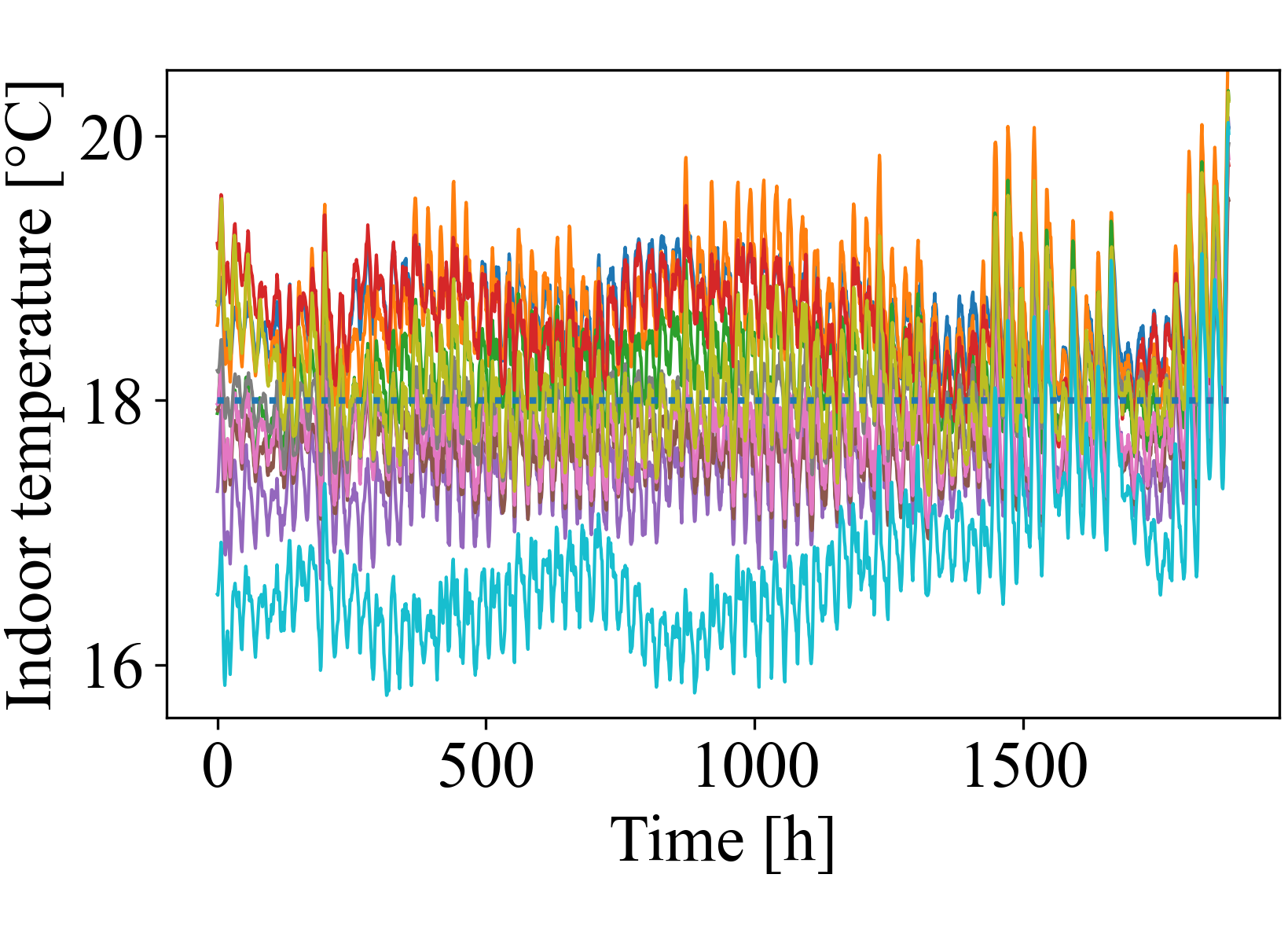}
}
\subfigure[]
{
\includegraphics[width=0.45\textwidth]{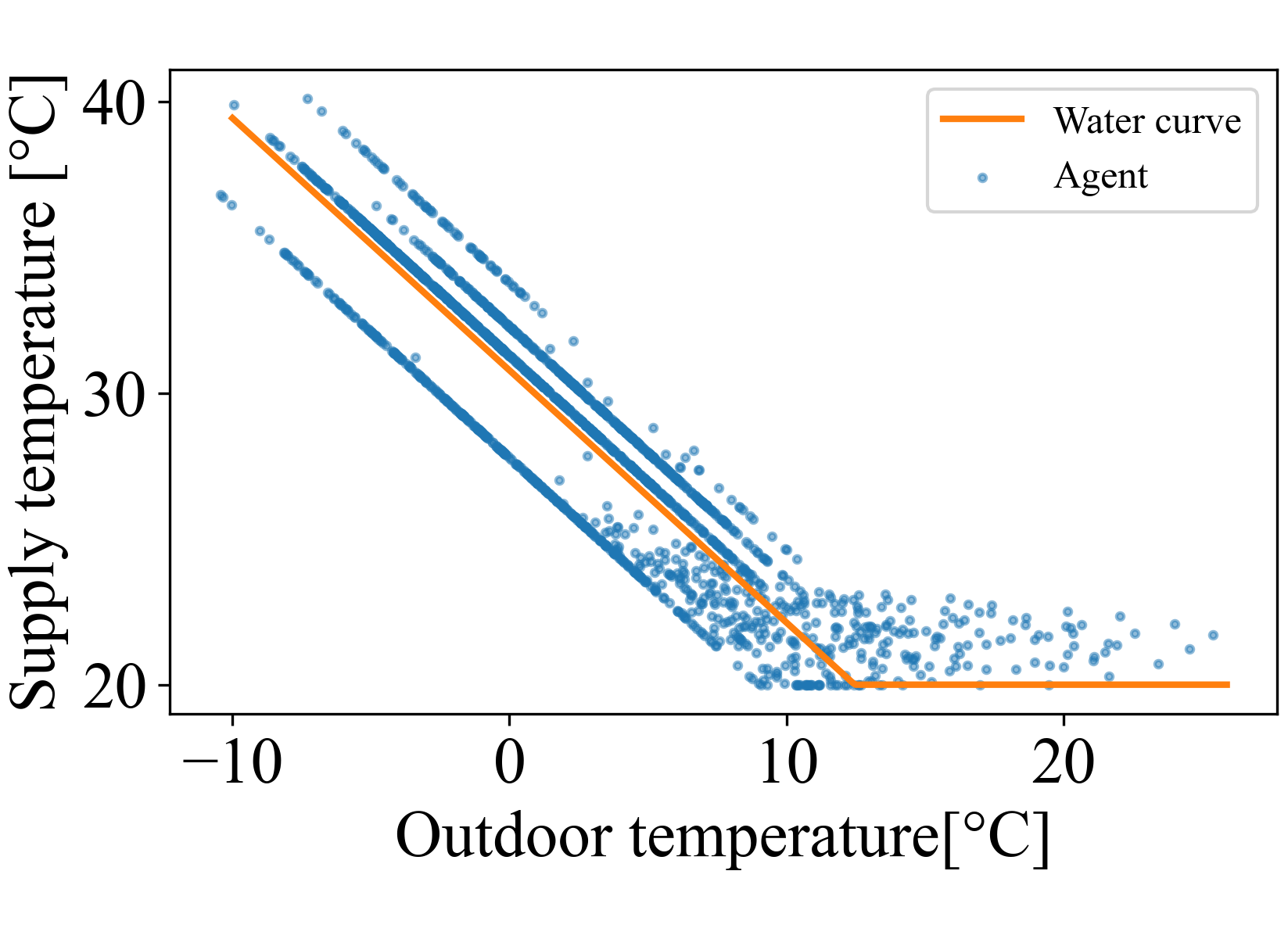}
}
\subfigure[]
{
\includegraphics[width=0.45\textwidth]{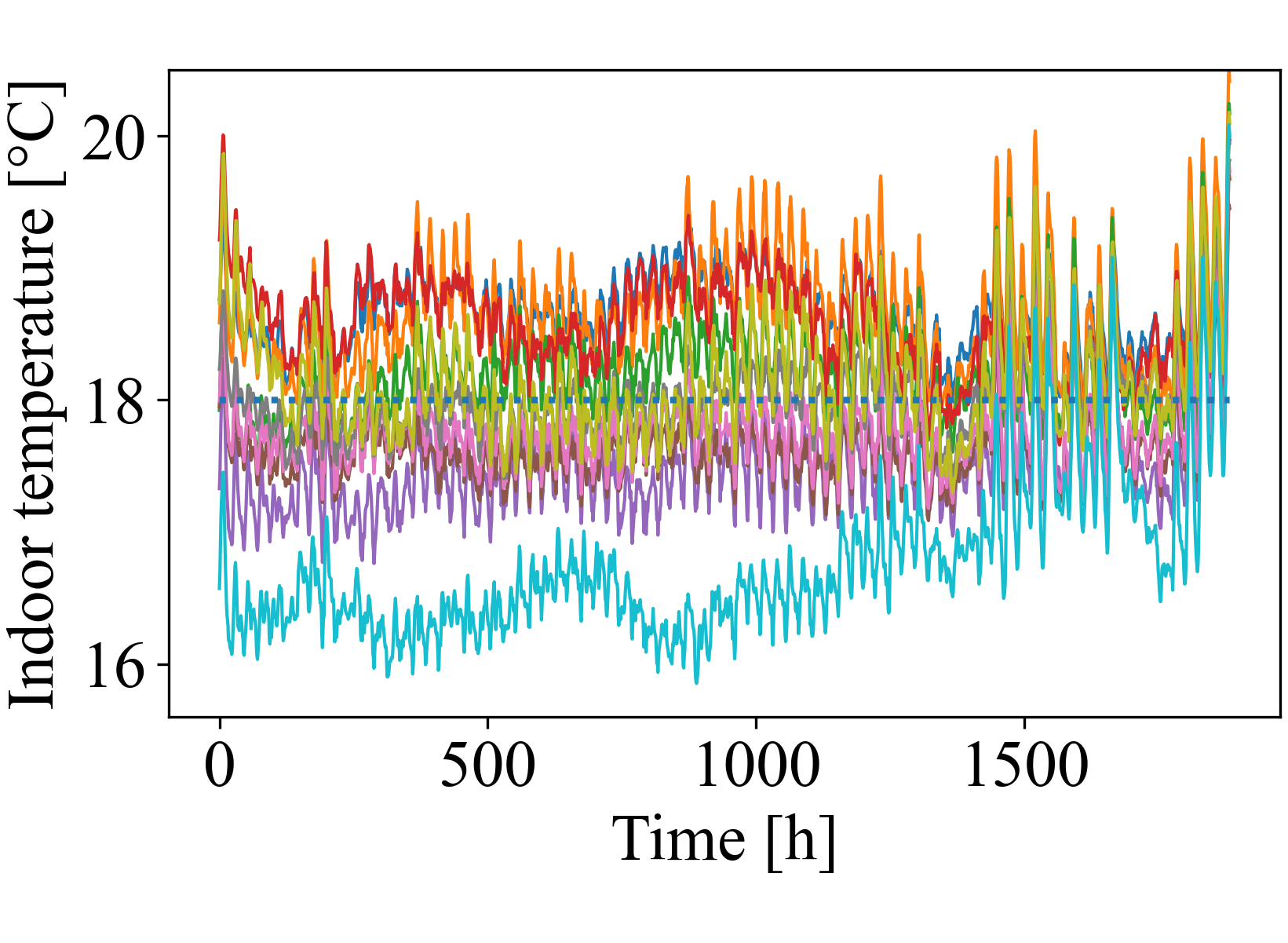}
}
\subfigure[]
{
\includegraphics[width=0.45\textwidth]{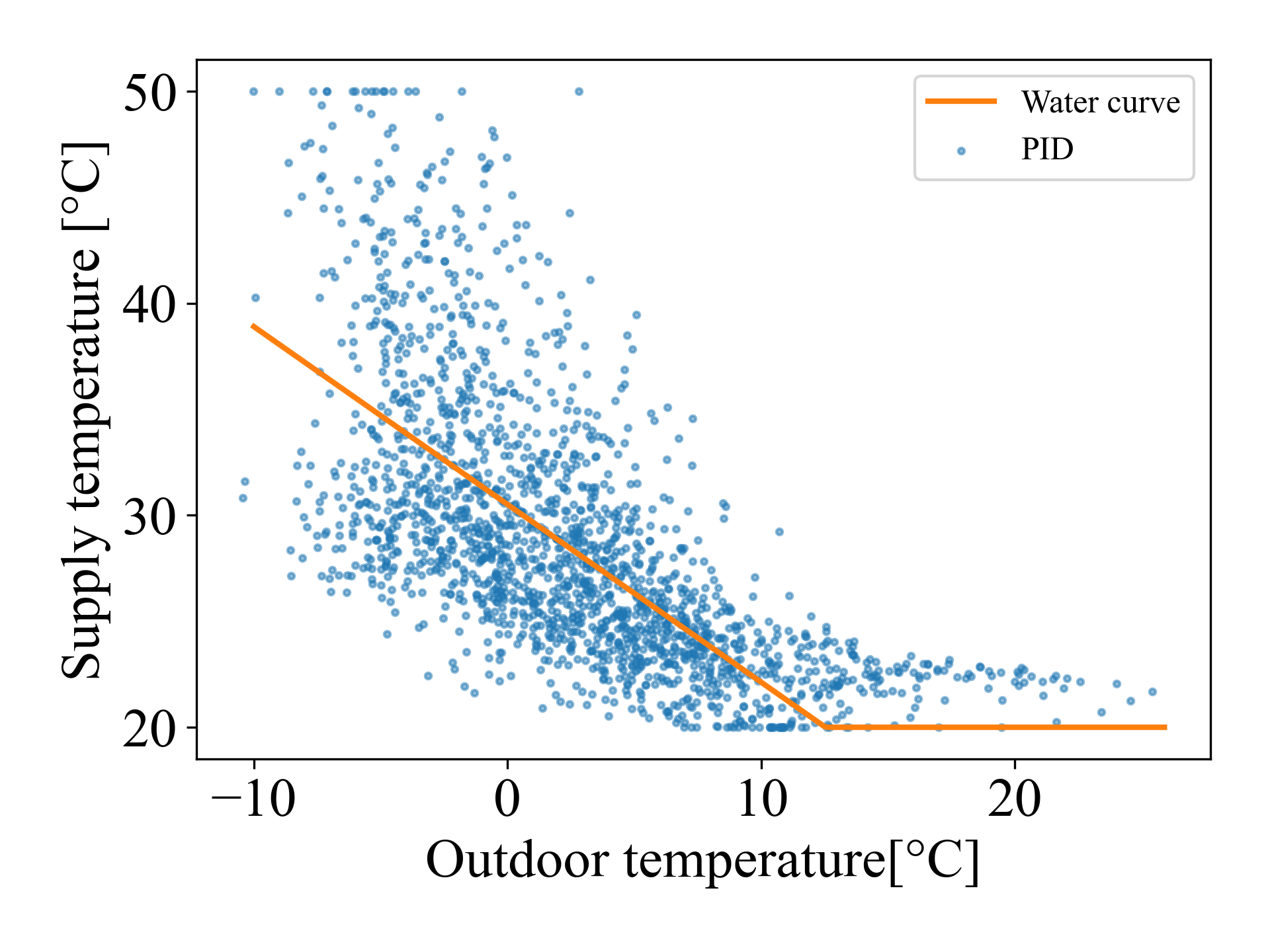}
}
\subfigure[]
{
\includegraphics[width=0.45\textwidth]{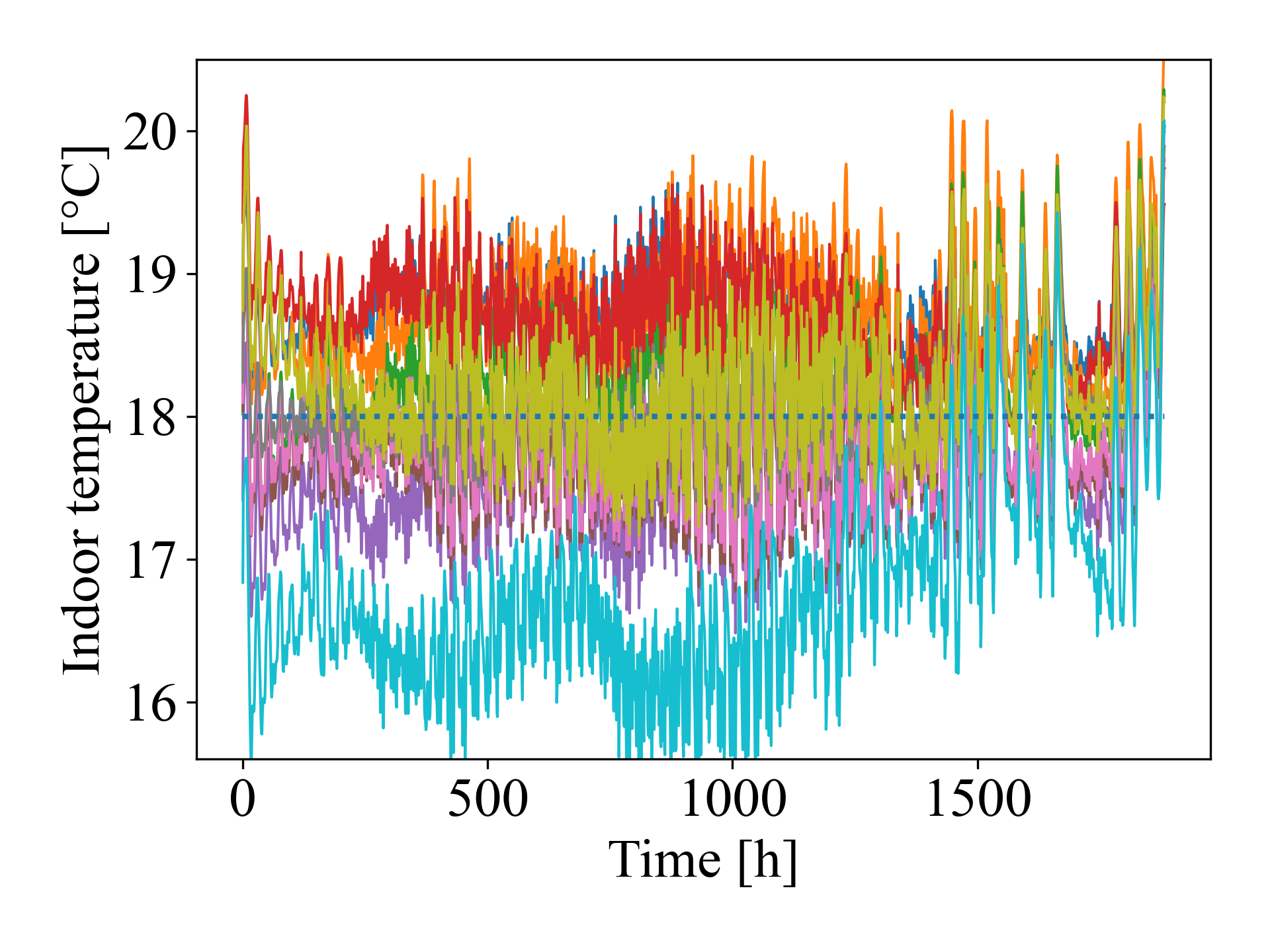}
}
\caption{
Control strategies of (a) Agent 1 (c) Agent 2 and (e) PID against the water curve, for multiple apartment and $\mathcal{T}\equiv18\celsius$.
Performances of (b) Agent 1, (d) Agent 2 and (f) PID in terms of $T_{in}$ for multiple apartments during the heating season.
Best viewed in colors.
}
\label{fig:res-tin-multi-apt2}
\end{figure}

Additional figures are provided for the experiments with multiple apartments:
Figure \ref*{fig:rewards} shows the reward of both agents during the 300 episodes of training and Figure \ref*{fig:res-tin-multi-apt2} shows the optimal control strategies and indoor temperatures for the two agents and the baseline.

\begin{figure}
\centering
\includegraphics[width=0.4\textwidth]{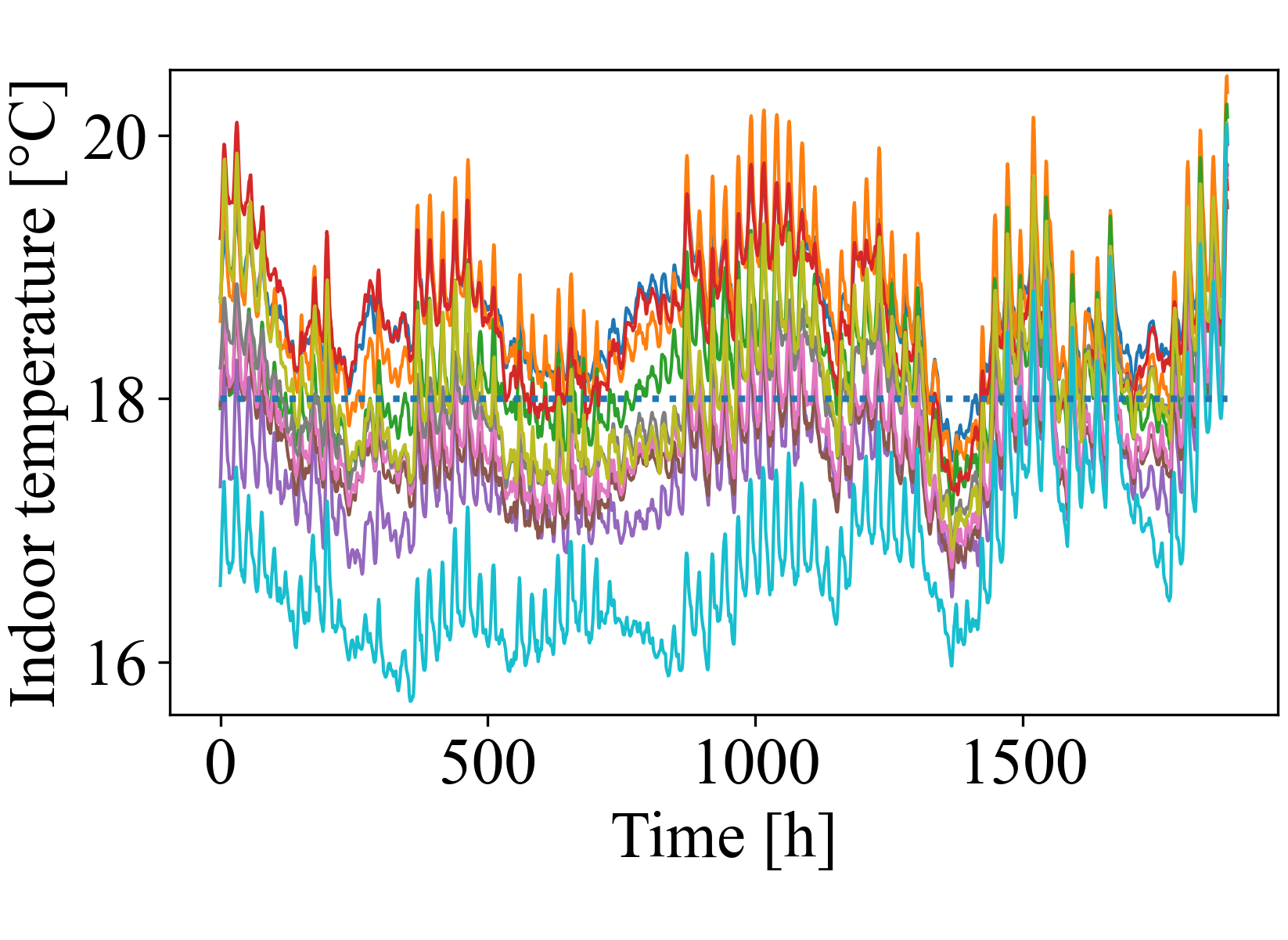}
\caption{Indoor temperature trajectories of the baseline control during the heating season.}
\end{figure}

\end{document}